\documentclass[%
aip, apl,
amsmath,amssymb,
 reprint,%
floatfix,
]{revtex4-1}
\usepackage{graphicx}
\usepackage{dcolumn}
\usepackage{bm}

\usepackage[utf8]{inputenc}
\usepackage[T1]{fontenc}
\usepackage{mathptmx}

\usepackage{siunitx}
\usepackage{amssymb}

\usepackage{blindtext}
\usepackage[dvipsnames]{xcolor}
\usepackage[ngerman,english]{babel} 
\newcommand{\DNi}[0]{D$_\text{Ni}$}
\newcommand{\DPd}[0]{D$_\text{Pd}$}
\newcommand{\DUV}[0]{D$_\text{UV}$}
\newcommand{\DAlOx}[0]{D$_\text{AlOx}$}

\newcommand{\Dref}[0]{D$_\text{ref}$}

\newcommand{\AlOx}[0]{Al$_2$O$_3$}

\begin{document}

	\title[Title]{
		Can surface-transfer doping and UV irradiation during annealing improve shallow implanted Nitrogen-Vacancy centers in diamond?
	}

	\author{N. J. Glaser}
	\author{G. Braunbeck}
	\author{O. Bienek}
	\author{I. D. Sharp}
	\author{F. Reinhard}
	\email{friedemann.reinhard@wsi.tum.de}
	\affiliation{ 
		Walter Schottky Institute and Department of Physics, Technical University of Munich, Am Coulombwall 4, 85748 Garching, Germany
	}

	\date{July 16, 2020} 
	
	\begin{abstract}
		It has been reported that conversion yield and coherence time of ion-implanted NV centers improve if the Fermi level is raised or lowered during the annealing step following implantation. 
		Here we investigate whether surface transfer doping and surface charging, by UV light, can be harnessed to induce this effect.
		We analyze the coherence times and the yield of NV centers created by ion implantation and annealing, applying various conditions during annealing. 
		Specifically, we study coating the diamond with nickel, palladium or aluminum oxide, to induce positive surface transfer doping, as well as annealing under UV illumination to trigger vacancy charging. 
        The metal coated diamonds display a two times higher formation yield than the other samples.
		The coherence time $T_2$ varies by less than a factor of two between the investigated samples.
		Both effects are weaker than previous reports, suggesting that stronger modifications of the band structure are necessary to find a pronounced effect.
		UV irradiation has no effect on yield and $T_2$ times.
	\end{abstract}
	\maketitle

	Nitrogen-vacancy (NV) centers have a great potential for quantum sensing. 
	The spin of individual NV centers can be controlled and read out at ambient conditions by microwave and optical excitation~\cite{gruber_1997}.
	Many applications have already been demonstrated, such as imaging magnetometry \cite{balasubramanian_2008, gross_2017} and the recording of NMR spectra from molecular-size samples \cite{staudacher_2013, mamin_2013}. 
	Furthermore, small local temperature differences \cite{neumann_2013, kucsko_2013} and the movements of mechanical resonators \cite{kolkowitz_2012} can be measured.
	
	For quantum sensing, close proximity of the NV center to the sample is crucial, which is only achieved for ``shallow'' NV centers located few nanometers beneath the diamond surface. A high control of the NV-surface distance can be achieved by implanting nitrogen atoms at low (keV) energies, followed by an annealing process to form NV centers. However, the resulting centers feature significantly reduced  $T_2$ times ($10-\SI{100}{\micro\second}$) and a low ($1\%$) conversion yield between implanted nitrogen atoms and NV centers.
	This has been attributed to the formation of divacancies and higher order complexes, which compete with the NV formation and can act as paramagnetic defects or charge traps\cite{favarodeoliveira_2017}.

	By introducing a boron doped sacrificial layer in proximity to the implanted nitrogen layer, during annealing, $T_2$ times and the shallow NV center yield could be improved
	\cite{favarodeoliveira_2017}. This phenomenon is explained by an increased hole density, leading to a positive charging of single vacancies, thus suppressing the formation of divacancies due to the repelling Coulomb forces.
	Recently, a significant improvement in the formation yield (from \SI{2}{\percent} to over \SI{60}{\percent}) was achieved by directly n-doping the diamond with co-implanted phosphorous, oxygen or sulfur \cite{luhmann_2019}. This success has been attributed to the same mechanism, except exploiting negatively charged single vacancies.
	
	Both diamond doping and removal of a sacrificial layer are technically challenging and hence inaccessible to many laboratories. 
	We therefore investigate whether similar effects can be achieved by technically simpler and noninvasive methods, in particular surface transfer doping and UV illumination.
	
	Surface transfer doping is a method to induce a thin doped layer close to the diamond surface. 
	It typically employs H-termination of the diamond to favor transfer of electrons across the surface into adsorbed electrolytes \cite{maier_2000}, molecules \cite{strobel_2004}, or metal electrodes \cite{tachibana_1992}, inducing p-doping close to the surface. 
	This technique has been previously employed to control the charge state of defects close to the surface \cite{hauf_2011, schreyvogel_2015}.
	
\begin{figure}[tb]
		\includegraphics[width=\linewidth,height=0.35\textheight,keepaspectratio=true]{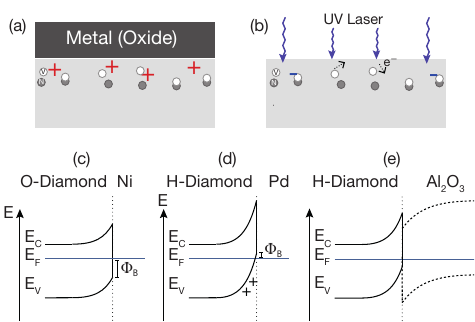}
		\caption{
			Schematic of the effects on the individual diamonds during annealing. (\textbf{a}) Three samples are covered with Ni, Pd or \AlOx{}, respectively, for surface doping. (\textbf{b}) \DUV{} was UV irradiated, to induce charge diffusion and bias the charge state of vacancies.
			(\textbf{c}) Nickel coating an O-terminated diamond leads to a band deformation by the formation of a Schottky contact, with the Schottky barrier height $\Phi_B$.
			(\textbf{d}) 
			Palladium coating raises the valence band of an H-terminated diamond over the Fermi level ($E_F$), resulting in a hole gas at the surface. 
			(\textbf{e})
			Diamond bands can be raised by a hetero junction with \AlOx{}, as the defect levels in diamond (N) and \AlOx{} (O$_i$) are leveled.
			\label{fig:methods} 
		}
		
	\end{figure}

	Here we aim to replicate the same effect 
	by forming a nickel junction on O-terminated diamond as well as a palladium junction or aluminum oxide junction on H-terminated diamond. In contrast to previous work, we cannot employ a bare H-terminated surface, as it would not be stable throughout the annealing. 
	Figure~\ref{fig:methods} summarizes the effects schematically.
	
	For a coating of O-terminated diamond with nickel (Fig.~\ref{fig:methods}a), the Fermi level of nickel is situated above the valence band, but below the Fermi level deep inside the diamond bulk. This latter level is expected to be set by the dominating impurity, substitutional nitrogen, to a level of $E_F = E_V + \SI{4.0}{\electronvolt}$ \cite{deak_2014}. 
	By forming a junction between both materials, electrons will therefore diffuse into the diamond, to form a thermal equilibrium. This leads to the formation of a band bending within the diamond. 
	The junction exhibits a Schottky barrier height $\Phi_B= E_F - E_V$, at the interface, between \SI{1.1}{\electronvolt} and \SI{1.7}{\electronvolt}  \cite{tsugawa_2010, vanderweide_1994}. 
	Hence, vacancies in close proximity to the surface will be forced into the $V^0$ state.
	Neutral vacancies are known to have a higher mobility in the diamond than $V^-$~\cite{davies_1992, deak_2014}, yet are believed  to be more prone to clustering.
	
	A H-terminated diamond surface induces an even stronger shift of the Fermi level, owing to the negative electron affinity $\chi=E_C-E_\text{vac} \approx -\SI{1.3}{\electronvolt}$ of H-terminated diamonds \cite{maier_2001}. A junction with the metal palladium will shift the diamond Fermi level near the surface into the valence band, forming a thermally stable Ohmic contact with $\Phi_B = E_F - E_V =  \SI{-0.15}{\electronvolt}$ \cite{wang_2015}. This will discharge vacancies close to the surface into a positive charge state ($V^+$, $V^{2+}$), possibly creating electrostatic repulsion between them \cite{favarodeoliveira_2017}. 
	
	A similar situation is expected for an interface between diamond and the insulator aluminum oxide \AlOx{}, where a p-doped hole gas has been observed \cite{kueck_2010, kasu_2012}, and has been explored for power electronics in recent experiments \cite{oi_2018}.   
	The exact spatial dependence of band bending depends on several unknown parameters, in particular the dominating species and concentration of defects in the aluminum oxide, as well as the concentration of defects at the interface. 
	Hence, no exact value for the barrier height can be given, making this case less well controlled than the metal contacts.
	
	As a possible noninvasive and simple method to charge lattice defects, we investigated the effect of laser irradiation during the annealing process. We irradiated the diamond with a wavelength of \SI{405}{\nano\meter} (\SI{3.06}{\electronvolt}), biasing the vacancies into the negative $V^-$ charge state.
	The selected irradiation energy is located above the threshold required for $V^0\rightarrow V^-$ conversion (reported to be \SI{2.88}{\electronvolt} \cite{neves_2001}) and below the threshold for the inverse process (\SI{3.15}{\electronvolt}\cite{neves_2001}).
	Ionizing other defects inside the diamond might lead to a further diffusion of charges influencing the diffusion of vacancies.

	For this study, four IIa electronic grade (100) diamonds from \textit{Element Six Ltd.}\ were used (size $\SI{2}{\milli\meter}\times\SI{2}{\milli\meter}\times\SI{0.5}{\milli\meter}$, $[\rm{N}]<\SI{5}{ppb}\, \hat{=}\, \SI{9 E14}{\per\centi\meter\cubed}$, roughness $\text{Ra}<\SI{5}{\nano\metre}$).

    \begin{table}[tb]
		\caption{Sample preparation details.}{\label{tab:preparation}}
		\begin{ruledtabular}
			\begin{tabular}{lcccc}
				& \DNi{} & \DPd{} & \DUV{} & \DAlOx{} \\
				\hline
				Applied Material & Nickel & Palladium &  & Al$_2$O$_3$  \\ \hline
				N-Implantation & $\checkmark$ & $\checkmark$ & $\checkmark$ & $\checkmark$ \\\hline
				Piranha cleaning & $\checkmark$ & $\checkmark$ & $\checkmark$ & $\checkmark$ \\\hline
				O-Plasma & $\checkmark$ & $\checkmark$ &   & $\checkmark$   \\\hline
				H-Plasma &   & $\checkmark$ &   & $\checkmark$  \\\hline
				Annealing &  \SI{830}{\celsius}   & \SI{830}{\celsius}  &  \SI{830}{\celsius}  + UV& \SI{830}{\celsius} \\\hline
				Aqua regia& $\checkmark$ & $\checkmark$ &  & \\\hline
				Hydrofloric acid&  &  &  & $\checkmark$  \\\hline
				3-acid-cleaning & $\checkmark$ & $\checkmark$ & $\checkmark$ & $\checkmark$ \\\hline
				O-Plasma &   & $\checkmark$ &   & $\checkmark$  \\
			\end{tabular}
		\end{ruledtabular}
	\end{table}

	The main processing steps to form the NV centers in the individual samples are summarized in table~\ref{tab:preparation}.
	The diamonds were acid cleaned (4 hours in a boiling H$_2$SO$_4$:HNO$_3$:HClO$_4$ mixture) and jointly implanted with $^{15}$N$^{+}$-ions (implantation energy \SI{5}{\kilo\electronvolt},  ion fluence \SI{5e9}{ions \per\centi\meter\squared}) at a \SI{7}{\degree} angle (\textit{Cutting Edge Ltd.}). We expect an implantation depth of \SI{10}{\nano\meter}~\cite{staudacher_2012, pezzagna_2010}.
	
	The diamond samples \DNi{}, \DPd{} and \DAlOx{} were oxygen terminated in an oxygen plasma (\textit{PS100-E, PVA TePla}, $p=\SI{1.4}{\milli\bar}$, power \SI{200}{\watt}) for \SI{300}{\second}.
	\DPd{} and \DAlOx{} have been further fully hydrogen terminated by a hydrogen plasma in a microwave reactor (Astex,
	\SI{15}{\minute} at \SI{700}{\watt} (\DPd{}) or \SI{750}{\watt} (\DAlOx{}), temperature \SI{700}{\celsius})\cite{dankerl_2009}.
	
	\SI{50}{\nano\meter} films of nickel or palladium were deposited onto \DNi{} and \DPd{} by electron beam evaporation ($p<\SI{E-6}{\milli\bar}$, deposition rate: \SI{1.1}{\angstrom\per\second}) on the O- or H-terminated diamonds, respectively.
	Atomic layer deposition was used to grow a \SI{10}{\nano\meter} thick layer of aluminum oxide onto \DAlOx{} ($T_\text{substrate} = \SI{200}{\celsius}$, p = \SI{0.25}{\milli\bar}), with successive cycles of trimethylaluminum and water exposure.
	The diamonds were annealed in a vacuum of $\sim\SI{E-6}{\milli\bar}$. After a \SI{60}{\minute}  heat up period, the diamonds were held for \SI{225}{\minute} at \SI{830}{\celsius}.
	\DUV{} was irradiated by a UV laser during the annealing process ($\lambda = \SI{405}{\nano\meter}$, laser power $\sim\SI{250}{\milli\watt}$, illuminated area $\sim \SI{4}{\milli\meter\squared}$).

	After annealing, the nickel and palladium layers were removed by an aqua regia solution. The aluminum oxide layer was removed by hydrofluoric (HF) acid. To remove the graphite layer resulting from the annealing, all diamonds were acid cleaned, as described above.
	The H-terminated diamonds \DPd{} and \DAlOx{} were finally O-terminated again.

	The measurements of the spin properties of the NV centers were performed with a home-built confocal microscope. 
    Excitation with a \SI{532}{\nano\meter} laser and imaging were performed through an oil-immersion objective ($\text{NA}=1.35$). The emitted photoluminescence was filtered by a \SI{650}{\nano\meter} long pass filter and measured by an avalanche photodiode. All measurements were performed at room temperature.
	
	ODMR, Rabi and $T_2$ measurements were performed at a resonance frequency of \SI{1.6}{\giga\hertz}, corresponding to a magnetic field $B = \SI{45}{\milli\tesla}$.
	Lorentzian curves were fitted to the data obtained by ODMR measurements.
	The $T_2$ time was measured by Hahn echo sequences with a rabi frequency of~$\approx \SI{9}{\mega\hertz}$.  The data points were fitted to an ESEEM curve involving a stretched exponential decay, considering collapses and revivals induced by the $^{13}$C bath.
	
	To relate the fluorescence properties and $T_2$ times of the samples, data of a diamond sample (denoted here as \Dref{}) that was prepared using the standard annealing procedure was used from a previous study\cite{braunbeck_2018}.

	\begin{figure}[tb]
		\includegraphics[width=\linewidth,height=0.7\textheight,keepaspectratio=true]{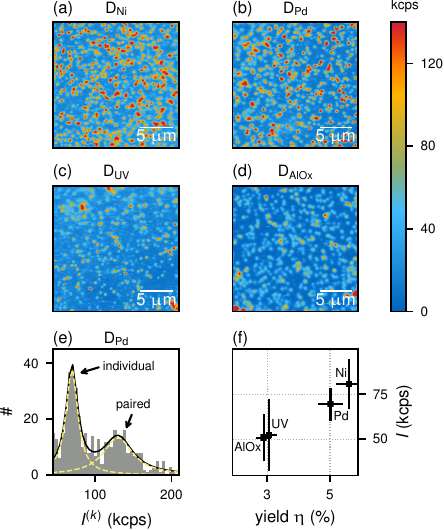}
		\caption{\label{fig:confocalscans}
			(\textbf{a}-\textbf{d}) Confocal fluorescence scans of the sampled regions.
			(\textbf{e}) Histogram of the fluorescence intensity I$^\text{(k)}$ for \DPd{}. A curve (black) consisting of two Lorentzians (yellow) was fitted to the histogram, distinguishing between individual and paired NV centers.
			(\textbf{f}) Mean fluorescence intensity $I$ and NV-center yield $\eta$ of the respective samples.}
	\end{figure}

	
	Figure~\ref{fig:confocalscans}a-d shows confocal fluorescence images of representative regions on the respective samples.
	We will analyze these images with regard to (i) fluorescence intensity of the individual NV centers, (ii) the density of NV centers, i.e. the $\text{N}\rightarrow \text{NV}$ conversion yield, and (iii) the background fluorescence intensity~$I_B$. 
	Luminescence peaks (NV centers) of the fluorescence image are identified by points, which are maxima within a $\SI{0.3}{\micro\meter} \times \SI{0.3}{\micro\meter}$ environment and a luminescence well above the background luminescence $I_B$. The fluorescence intensity $I^{(k)}$ of individual NV centers is calculated by subtracting the background luminescence  from the respective peak luminescence $I_{max}$ according to  $I^{(k)}=I^{(k)}_{max}-I_B$.
	Looking at the histogram of the luminescence intensities $I^{(k)}$ (Fig. \ref{fig:confocalscans}e) we find a bimodal distribution. As we can only distinguish NV centers with a separation greater than the confocal volume, we attribute the peaks to individual (low-countrate) and pairs of (high-countrate) NV centers within the acquired resolution.
	This distribution was fitted to two Lorentzian peaks. 
	In order to calculate the overall yield $\eta$, a linear model was established, relating $N / N_0$ and $a_2 = A_2 / (A_1 + A_2)$ using synthetic data of randomly distributed NV centers, with the true number of NV centers $N$, the number of identified peaks $N_0$ and the respective integrated areas of the Lorentzian curves $A_i$.
	With the assumption that the first peak is only attributed to single NV centers, we determine the mean fluorescence intensity $I$ of individual NV centers by the center of this peak, ignoring data points from the second peak.
	The result of this analysis is presented in table~\ref{tab:fluorescence} and leads to the following results:

	\begin{table}[tb]
		\caption{Fluorescence properties of the NV centers. Mean intensity of individual NV centers $I$, background fluorescence~$I_B$, NV conversion yield $\eta$ and ratio $R$ of the counted paired NV centers in the measurements to the number expected by simulations.
		\label{tab:fluorescence}}
		\begin{ruledtabular}
			\begin{tabular}{l|rrrr}
				&  $I$ (kcps) & $I_B$ (kcps) & $\eta$ ($\%$) & $R$ (1) \\ \hline
				\DNi{}   & $81 \pm 14$ &         15.3 &        $5.61$ &  $1.0$ \\
				\DPd{}   &  $69 \pm 9$ &          8.7 &        $5.01$ &  $1.6$ \\
				\DUV{}   & $52 \pm 20$ &         10.2 &        $3.05$ &  $0.5$ \\
				\DAlOx{} & $51 \pm 13$ &          1.9 &        $2.89$ &  $1.0$ \\
				\Dref{}  & $223 \pm 65$ &          8.5 &        $2.47$ &  $0.5$
			\end{tabular}
		\end{ruledtabular}
	\end{table}
	
	The fluorescence intensity of individual NV centers is higher for the samples annealed under a metal coating (Ni and Pd) than for the samples annealed under \AlOx{} coating and UV illumination. This hints towards a higher stability of the $NV^-$ charge state in the metal-coated samples. Since our detection optics are selective for $NV^-$, discharging into the neutral $NV^0$ state would manifest itself as a reduction of fluorescence intensity for this NV center. \par
	
	The yield $\eta$ of NV center creation follows a similar trend. $\eta$ is given by the ratio of the density of formed NV centers to the density of implanted nitrogen atoms.
	In relation to \DUV{} and \DAlOx{} the yield is increased by \SI{85}{\percent} for \DNi{} and by \SI{70}{\percent} for \DPd{} to $\eta > \SI{5}{\percent}$, as shown in Fig.~\ref{fig:confocalscans}f.
	Thus, by discharging V$^-$, an increased NV center yield is seen within our samples.
	
	The formation of paired NV centers seems to be correlated to the charge state of vacancies. The ratio $R=a_{2} / a_{2,\text{synth}}$ relates the amount of paired NV centers in the measured samples (Fig. \ref{fig:confocalscans}f) to the expected number from simulations. 
	While annealing, the NV centers are dominantly in the negative charge state NV$^-$.
	However, the vacancies may change their charge state due to surface doping, as they are in closer proximity to the surface. 
	\DPd{}, promoting positively charged vacancies (V$^{+}$), has a \SI{60}{\percent} 
	increase of pairs compared to calculations (Tab. \ref{tab:fluorescence}). Vacancies might be attracted by existing NV$^-$ centers, enhancing the formation of new NV centers in close proximity to other NV$^-$ centers.
	In \DNi{} and \AlOx{} no deviation from the expected value is observed, hence a random formation is achieved, with vacancies being in the neutral state (V$^0$).
	The UV irradiated and the reference diamond have $R$ values well below 1, suggesting repulsion for negatively charged V$^-$.\par
	
	We observe that the type of termination {\it during} the annealing process does not appear to influence the brightness of the NV centers, as long as the H-terminated samples are O-terminated again {\it after} annealing. 
	After the annealing and acid cleaning process, 
	the NV center luminescence intensities in \DPd{} and \DAlOx{} were greatly reduced. In combination with a missing response on the microwave signal this suggests quenching of the NV centers into the neutral NV$^0$ state. After the final O-Plasma treatment, the luminescence intensity increased to a similar level as found for non H-plasma treated diamonds, recovering the NV$^-$ state. 
	
	The background fluorescence intensity $I_B$ indicates the concentration of optically active lattice defects within the diamond and surface contamination. This value varies among the samples between \SI{1.9}{kcps} (\DAlOx{}) and \SI{15.3}{kcps} (\DNi{}).
	The particularly low background fluorescence of \DAlOx{} might show a reduced defect concentration. Alternatively,  the HF acid cleaning, which is necessary
	to remove the \AlOx{} layer, might have removed further surface contamination, resulting in a reduced fluorescence of the nearby surface.

	\begin{figure}[tb]
		\includegraphics[width=\linewidth,height=0.45\textheight,keepaspectratio=true]{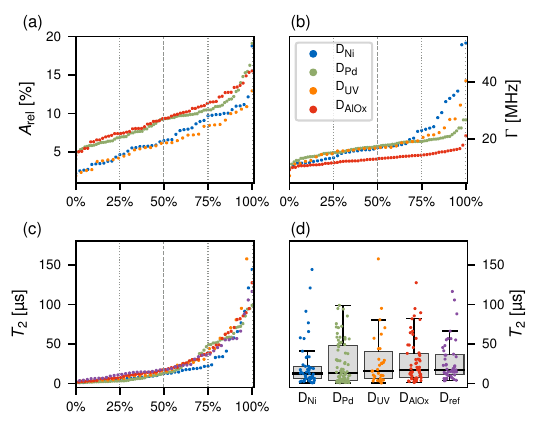}
		\caption{\label{fig:spinprops}
			Spin properties of the NV-centers. 
			Distribution of (\textbf{a}) the relative amplitude $A_\text{rel}$ and (\textbf{b}) the FWHM $\Gamma$ of the ODMR dips, as well as (\textbf{c}) $T_2$ times of individual NV centers.
			The measured values are uniformly distributed with rising values along the abscissa. 
            Box plots of the measured $T_2$ times are shown in (\textbf{d}) for the respective diamonds.
			The gray box is limited by the 1st and 3rd quartile, thus including \SI{50}{\percent} of the data points. The black bar indicates the median of the distribution.
            \Dref{} denotes the $T_2$ times of a previous study\cite{braunbeck_2018} with NV centers formed by the standard annealing procedure.
            }
	\end{figure}
	
	Figure~\ref{fig:spinprops} shows the distribution of relevant spin properties, where each dot represents an NV center: 
	(a) the contrast of the optically detected magnetic resonance (ODMR)  signal $A_\text{rel}$, (b) the ODMR linewidth $\Gamma = (\pi T_2^*)^{-1}$, and 
	(c,d) the spin coherence times $T_2$. This analysis generates the following insights:
	
	Fig.~\ref{fig:spinprops}a illustrates that the observed ODMR contrast varies between the investigated samples.
	The NV centers of 
	\DAlOx{} and \DPd{} show the highest ODMR contrast, which is about  50\% larger compared to \DNi{} and \DUV{}. This is different from the behaviour observed in fluorescence properties, where \DNi{} and \DPd{} showed a comparable behaviour. 
	Which seems to be due to the fact that the ODMR contrast is increased by several factors, most prominently a high stability of the $NV^-$ state (\DPd{}) and a narrow ODMR linewidth (\DAlOx{}). 

	The linewidth $\Gamma$ of the ODMR spectra (Fig.~\ref{fig:spinprops}b) is representative for the spin dephasing time $T_2^*$.
	The linewidth $\Gamma$ is smallest for \DAlOx{}, corresponding to the longest $T_2^*$ times.  
	In addition, \DAlOx{} and \DPd{} show the narrowest distribution of $\Gamma$. 
	This suggests that the presence of a hole gas during annealing (or the oxygen plasma for removal of the hydrogen termination) improves $T_2^*$. 
	Roughly \SI{25}{\percent} of the NV centers in \DNi{} have a substantially increased linewidth~$\Gamma$.
	
	
    The distributions of the $T_2$ times (Fig.~\ref{fig:spinprops}c,d) show that the median values (corresponding to the $T_2$ values at 50\%  in Fig.~\ref{fig:spinprops}c and indicated by black bars in Fig.~\ref{fig:spinprops}d)  vary only in a small range between \SI{12}{\micro\second} (\DNi{}) and \SI{17}{\micro\second} (\DAlOx{}, which also had the longest $T_2^*$). 
	However, the tails of the distributions vary between the samples. For example, the upper quartile of the respective samples is moderately increased for \DPd{} and halved for \DNi{} in comparison to the three other samples.
	Both, the relatively large ODMR line broadening $\Gamma$ and the relatively narrow upper quartile of the \DNi{} sample
	suggest some source of magnetic impurities. We speculate that even after the final etching process, a small concentration of residual nickel atoms with a strong magnetic moment could remain on the diamond surface, reducing the dephasing and coherence time of nearby NV centers.
	Nevertheless, as can be seen in Fig.~\ref{fig:spinprops}c, in all samples the top \SI{10}{\percent} of NV centers  have similarly long $T_2$ values in the range between \SI{59}{\micro\second} and \SI{78}{\micro\second}.
	This shows that a small population of comparably robust NV centers exists in all investigated samples, which  might be intrinsic to the diamond or lie deeper within the bulk. 

	In summary, we investigated UV illumination and surface transfer doping as tools to temporarily dope a diamond during the formation of implanted NV centers by
	annealing. 
    As its most salient conclusion, our study reveals that these tools have a weaker effect on the conversion yield and the spin properties than doping by growth or co-implantation, however the effect is non-negligible.
    Conversion yields vary by a factor of two between samples, but have been demonstrated to vary by a full order of magnitude for co-implantation, using however less shallow NV centers, hence being more aided by the bulk properties. The coherence time $T_2$ also varies by less than a factor of two between our samples, but has been reported to improve by a full order of magnitude by temporary Boron doping.
    Additionally we have to note that the low magnitude of the effects and the small number of samples generate some sensitivity to systematic errors, such as different properties of the underlying diamond substrates, implantation regions or hydrogen contents.
	Nonetheless, we draw the following conclusions:

	Coating the diamond by metals (nickel or palladium) results in an increased yield by a factor of two compared to the non metal coated samples, and partially induces enhanced paired formation of NV centers (palladium). It also improves the NV center fluorescence intensity, hinting towards a higher stability of the NV$^-$ charge state. 
	The spin dephasing time $T_2^*$ improves significantly for an \AlOx{} coating and slightly for a palladium coating, evidenced by a narrow ODMR linewidth, suggesting an effect of transfer doping, HF-treatment or the final O-plasma exposure. 
	In contrast, the coherence time $T_2$, arguably the most important figure of merit, is affected least strongly by our treatments. We observe a slight improvement for palladium coating and a clear reduction for nickel coating. The latter could be due to residual magnetic adsorbents or clustering of vacancies. 
	Finally, UV irradiation has no effect on $T_2$ times and seems to degrade fluorescence intensity as well as $T_2^*$.
	\begin{acknowledgments}
		This work has been supported by the European Union (Horizon 2020 research and innovation programme, grant
        820394 (ASTERIQS)) and the Deutsche Forschungsgemeinschaft 
        (German Excellence Strategy -- EXC-2111 -- 390814868 and Emmy Noether grant RE3606/1-1). We thank Patrick Simon for hydrogen terminating samples of this study.
	\end{acknowledgments}
	
	\section*{Data Availability}
	Data available on request from the authors

	\bibliography{../../../../../../ZoteroBibWin}
	
\end{document}